\documentclass[10pt, journal, colorlinks]{IEEEtran}
\usepackage{natbib}
\bibpunct{(}{)}{;}{a}{}{,}
\usepackage[pdftex]{graphicx}
\usepackage{txfonts}
\begin{document}

\newcommand\plotone[1]{%
 \leavevmode 
 \includegraphics[width={\columnwidth}]{#1}%
}%

\newcommand\plottwo[2]{%
 \centering 
 \leavevmode 
 \includegraphics[width={0.50\columnwidth}]{#1}%
 \includegraphics[width={0.50\columnwidth}]{#2}%
}%

\newcommand{\plotthree}[3]{ 
 \centering 
 \leavevmode 
 \includegraphics[width={0.32\columnwidth}]{#1}
 \includegraphics[width={0.32\columnwidth}]{#2}
 \includegraphics[width={0.32\columnwidth}]{#3}
}

\newcommand{\Hogbom}{{H\"ogbom}}
\newcommand{\pasp}{{Publ. Astr. Soc. Pacific}}
\newcommand{\aj}{{Astron. Journal}}
\newcommand{\apj}{{Astrophys. Journal}}
\newcommand{\aap}{{Astron. \& Astrophys.}}
\newcommand{\aaps}{{Astron. \& Astrophys. Suppl.}}
\newcommand{\nat}{{Nature}}
\newcommand{\mnras}{{Mon. Not. Roy. Astr. Soc.}}

\title{Multi-Scale CLEAN deconvolution of radio synthesis images} 

\author {T.J. Cornwell

\thanks{T.~J.~Cornwell is with the Australia Telescope National Facility, Epping, NSW,
  Australia}}

\maketitle

\begin{abstract}
  Radio synthesis imaging is dependent upon deconvolution algorithms
  to counteract the sparse sampling of the Fourier plane. These
  deconvolution algorithms find an estimate of the true sky brightness
  from the necessarily incomplete sampled visibility data. The most
  widely used radio synthesis deconvolution method is the CLEAN
  algorithm of \Hogbom. This algorithm works extremely well for
  collections of point sources and surprisingly well for extended
  objects. However, the performance for extended objects can be
  improved by adopting a multi-scale approach. We describe and
  demonstrate a conceptually simple and algorithmically
  straightforward extension to CLEAN that models the sky brightness by
  the summation of components of emission having different size
  scales. While previous multiscale algorithms work sequentially on
  decreasing scale sizes, our algorithm works simultaneously on a
  range of specified scales. Applications to both real and simulated
  data sets are given.
\end{abstract}

\date{Received date; accepted date}
\markboth{Multi-Scale CLEAN}
\maketitle


\newcommand{\AIPSpp}{{\tt AIPS++/CASA}}
\newcommand{\Ff}[1]{   {\cal F}\left(#1\right)}
\newcommand{\Fi}[1]{   {\cal F}^{-1}\left(#1\right)}

\section{Introduction}

Radio synthesis imaging of astronomical sources was revolutionized by
the invention of the CLEAN algorithm \citep{Hogbom1974}. This simple
algorithm enabled synthesis imaging of complex objects even with
relatively poor Fourier plane coverage, such as occurs with partial
earth rotation synthesis or with arrays composed of small numbers of
antennas \citep[see][]{Thompsonetal2001}.

The algorithm is motivated by the observation that in the image formed
by simple Fourier inversion of the sampled visibility data (the
``dirty'' image), each point on the sky is represented by a suitably
scaled and centered point spread function. Hence one can find the
brightest point source by simply performing a cross-correlation of the
dirty image with the point spread function. Removing the effects of
the brightest point then enables one to find the next brightest, and
so on. This simple algorithm works very well for collections of point
sources, and surprisingly well for extended objects. Convergence is
slow for extended objects (as might be expected since one is trying to
identify a potentially large number of pixels one by one), and
instabilities occur \citep{Cornwell1983}. For this reason, other
deconvolution algorithms such as the Maximum Entropy Method 
\citep{GullDaniell1978,CornwellEvans1985} are often preferred. 
However, the CLEAN algorithm remains attractive for many purposes so
the natural question arises as to how the basic algorithm can be
improved in convergence and stability while maintaining the advantages
of simplicity and noise behavior.

There are many different deconvolution algorithms in use in various
scientific domains but one thread common to many of the best
algorithms is the use of the multi-scale approach \citep{Starcketal2002}. 
In multi-scale methods, the object to be recovered is
modelled as being composed of various different scale sizes. The
reconstruction algorithm then has the task of estimating the strengths
of the various scales, instead of estimating the strength of
pixels. Thus the number of degrees of freedom in the reconstruction
can be significantly reduced with a concomittant increase in various
measures of performance, such as robustness, stability, and signal to
noise. Examples of multi-scale methods are:

\begin{itemize}
\item {\bf Multi-Resolution CLEAN}: \cite{WakkerSchwarz1988} developed
  a simple strategy for running the CLEAN algorithm emphasizing broad
  emission first and then finer and finer resolution. The dirty image
  and point spread function are smoothed and decimated to emphasize
  the broad emission. The image resulting from CLEANing this dirty
  image is then used as an initial model for a CLEAN deconvolution of
  the full resolution image.
\item {\bf Multi-scale Maximum Entropy}: \cite{Weir1992}, and
  \cite{Bontekoeetal1994} noted that the performance of Maximum
  Entropy deconvolution could be improved by decomposing the image to
  be estimated into several channels of different resolutions. A
  hierarchy of scale sizes is specified and an image reconstructed by
  estimating pixels in the combined space such that the convolution
  equation is satisfied.
\item {\bf Wavelets} Numerous authors have described the virtues of
  wavelet analysis and its application to deconvolution. The recent
  textbook by \cite{Starcketal1998} provides an excellent summary and
  includes discussion of the connections between Multi-Resolution
  CLEAN and wavelet analysis. Various authors have described an
  extension of the Maximum Entropy Method to wavelets as basis
  functions \cite{PantinStarck1996,Starcketal2001,Maisingeretal2004}.
\item {\bf Pixons} \cite{PuetterPina1993} developed a method for
  estimating not only pixel strengths but an associated scale
  size. The combination of strength and scale size, they dubbed a
  ``pixon''. The Pixon method has been extended considerably, and the
  original algorithm drastically improved
  \citep{Dixonetal1996,PuetterYahil1999}. Performance is extremely
  good, especially as measured by the statistical whiteness of the
  residuals. However, there has been no published success in applying
  the algorithm to synthesis observations because a key assumption,
  that the PSF is compact, does not hold for Fourier synthesis.
\item {\bf Adaptive Scale Pixels} \cite{BhatnagarCornwell2004}
  developed a method for fitting extended components during a
  deconvolution process. This has good deconvolution performance but
  is computationally expensive.
\end{itemize}

Thus multi-scale methods have demonstrated advantages in
deconvolution, encouraging further development of multi-scale
algorithms. This conclusion is supported by the explosion of interest
and technical advances in ``compressive sampling''
\cite{HauptAndNowak2006a, HauptAndNowak2006b, BobinEtAl2008,
  BaraniukEtAl2008}. Compressive Sampling (CS) theory shows that under
quite general conditions, a sparse signal can be reconstructed from a
relatively small number of random projections. This is reassuring for
radio astronomers since deconvolution of radio synthesis observations
has been the norm for about thirty years.

The recent growth of work in compressive sampling algorithms holds
great promise for new and efficient algorithms for deconvolution of
radio synthesis observations. However, here we concentrate on a simple
extension of the CLEAN algorithm. We describe a conceptually and
algorithmically straightforward multi-scale generalization of the
CLEAN method that improves convergence and stability (as well as some
other properties). This retains the pattern matching motivation of the
\Hogbom\ algorithm but extends it to encompass extended emission as well
as point sources. Unlike the Multi-Resolution CLEAN and Wavelet CLEAN,
this algorithm selects among scales considered simultaneously rather
than sequentially. We present some background in the next section, our
multi-scale algorithm in section 3, some demonstrations and
comparisons to other algorithms in section 4, and summarize our work
in section 5.

\section{Background}

Radio synthesis arrays image the radio sky not through a single large
physical aperture but by synthesising a virtual aperture of equivalent
size and angular resolution. This is done by invoking the van
Cittert-Zernike theorem relating the spatial coherence function (or
``visibility'') of the electric field and the sky brightness
function  \citep[see][]{Thompsonetal2001}.

A given pair of antennas, with baseline vector $u,v,w$ (as seen from the source) 
measures a single Fourier component of the sky brightness $I$.

\begin{displaymath}
V(u,v) = \int I(x,y) e^{2\pi j (u x+v y)} dx dy
\end{displaymath}
Given complete sampling of the Fourier space, the sky brightness may be obtained by
Fourier inversion of noise free observations:

\begin{displaymath}
I(x,y) = \int V(u,v) e^{-2\pi j (u x+v y)} du dv
\end{displaymath}
For a real array, the true, completely sampled, visibility function
$V$ is not available and we have only noisy samples of the visibility
function at discrete locations in the Fourier plane. Ignoring for the
moment the effects of noise, we can represent this by replacing $V$ by
the ``sampled'' visibility function $S(u,v) V(u,v)$, where the
sampling function $S(u,v)$ is:

\begin{displaymath}
S(u,v) = \sum_k w_k \delta(u-u_k) \delta(v-v_k)
\end{displaymath}
Inserting this into the Fourier inverse, we then obtain the {\em Dirty} image:

\begin{displaymath}
I^D = \Fi{S V}
\end{displaymath}
Applying the convolution theorem from Fourier transform theory, we find
that the dirty image is the convolution of the true image $I$ with a
``Dirty'' beam $B$:

\begin{displaymath}
I^D = B * I
\end{displaymath}
where the Dirty Beam or Point Spread Function $B=\Fi{S}$ is
given by:

\begin{displaymath}
B(x,y) = \sum_k \cos\left[2\pi \left(u_k x+v_k y\right)\right] w_k
\end{displaymath}
Thus the deconvolution problem is to solve for $I$ from knowledge of
the dirty image $I^D$, and the point spread function $B$.

The CLEAN algorithm finds a solution to the convolution equation by
positing a model for the true sky brightness which is a collection of
point sources.

\begin{displaymath}
I^C = \sum_q I_q \delta (x-x_q) \delta (y-y_q)
\end{displaymath}
The key aspect of the CLEAN algorithm is the way that it solves
iteratively for the positions and strengths of the CLEAN
components. Defining:

\begin{displaymath}
I^C(m) = \sum_{q=1}^{m} I_q \delta (x-x_q) \delta (y-y_q)
\end{displaymath}
The values for $I_n,x_n,y_n$ are found by locating the
peak in the residual image:

\begin{displaymath}
I^R(n) = I^D - B * I^C(n-1)
\end{displaymath}
In fact, a least squares fit for $I_n,x_n,y_n$ dictates that one find
the peak in $B * I^R(n)$. However, the difference between these two is
often minor and is nearly always neglected. On finding the $n$th
component, the residual image is simply updated by subtracting a
suitably scaled and centered copy of the point spread function. Hence
the main work in the algorithm is (a) finding the location of the peak
residual, and (b) shifting, scaling, and subtracting the point spread
function. Performance of the deconvolution can be improved by
restricting the search to a ``CLEAN window'' wherein the image
brightness is known to be non-zero. Such support constraints aid
deconvolution in general by restricting the range of possible
solutions.

The \Hogbom\ CLEAN algorithm is therefore simple to understand and
implement. For small images, the algorithm is very fast. For larger
images, \cite{Clark1980} developed a faster algorithm that uses a
support-limited approximation to the point spread function, followed
by FFT-base convolution using the full point spread function.

\cite{Schwarz1978} has analyzed the \Hogbom\ CLEAN algorithm in detail
and has developed a convergence proof. CLEAN is a greedy algorithm: it
consists of a sequence of iterations, and within each iteration, the
choice made depends only on information available at the iteration
\citep[see][for a discussion of greedy algorithms.]{Cormenetal2001}
There is no guarantee that the result of the CLEAN algorithm is
globally optimal in any particular sense, although there is a
conjecture that the image L1 norm is minimized
\citep{MarshRichardson1987}. \cite{Schwardt2008} has developed the
interpretation of \Hogbom\ Clean in Compressive Sampling language as a
matching pursuit algorithm calculating residual vectors in image
space.

\section{The Multi-Scale CLEAN algorithm}

The \Hogbom\ CLEAN algorithm performs surprisingly well given that it
chooses to model extended emission by point sources (Dirac delta
functions). Our extension is to use extended components to model the
source. Our model is that that the true sky brightness is a summation
of appropriately scaled and centered extended components. Denoting the
component width by a single parameter $\alpha_q$, we have that the
model is:

\begin{displaymath}
I^M = \sum_q I_q m(x-x_q, y-y_q, \alpha_q)
\end{displaymath}
Following the greedy strategy used in the CLEAN algorithm, we search for the $n$th
component by looking for a peak in:

\begin{displaymath}
I^R(n) = I^D - B * I^M(n-1)
\end{displaymath}
where the search is over $I_q, x_q, y_q, \alpha_q$. In practice, the
search over the non-discrete axis $\alpha_q$ could be quite time
consuming. However, we have found that searching only a few
well-chosen scale sizes works well in many cases.

\begin{table}
\caption{Multi-Scale algorithm}
\begin{center}\framebox{
\begin{minipage}{0.95\columnwidth}
  {\bf Initialize}
  {\begin{itemize}
    \item Model: $I^M=0$
    \item Residual Image: $I^R=I^D$
    \item For each scale $\alpha_q$
      {\begin{itemize}
          \item calculate scale-convolved residual $I^R_{\alpha}=m(\alpha)*I^R$
          \item calculate scale bias $S(\alpha)=1-0.6*\alpha/\alpha_{\rm max}$
          \end{itemize}}
    \item For each pair of scales $\alpha_p, \alpha_q$, calculate cross term: $B*m(\alpha_p)*m(\alpha_q)$
    \end{itemize}}
  {\bf Repeat}
  {\begin{enumerate}
    \item For each scale, find strength and location of peak residual
    \item Choose scale with maximum residual, after multiplying by scale-dependent bias terms,
    \item Add this component to current model, scaled by loop gain
    \item Update all residual images using precomputed terms
    \end{enumerate}
  }
  {\bf Until}
  {\begin{description}
      \item[Either] $max(I^R_{\alpha}) <$ threshold
    \item[Or] Maximum number of components identified
  \end{description}}
  {\bf Finalize}
  {\begin{itemize}
      \item Convolve current model by clean beam: $B_G*I^M$
      \item Add residuals to get restored image: $B_G*I^M+I^R$
   \end{itemize}}
\end{minipage}
}
\end{center}
\end{table}

In deciding which scale to select, we must apply a bias towards
smaller scales.  To understand why, consider a source which is
dominated by a point source but which has a very small amount of
extended emission.  If we don't bias towards smaller scales, the point
source plus the small amount of extended emission will result in
selecting a large scale to represent the emission.  The residuals,
however, will show the point source nearly untouched, and a large
negative bowl around the point source, and it will take many
iterations to clean up this mess.  In any one iteration, we evaluate
the maximum residual for a set of scale sizes, and choose that scale
size with the maximum {\em adjusted} residual. We have found the following
relation to work well:

\begin{displaymath}
S(\alpha)=1-0.6 \alpha/\alpha_{\rm max}
\end{displaymath}
In any iteration, the component selected is thus located at the peak
with the appropriate scale size. The magnitude of the component is
given by the loop gain times the peak.  Unlike delta-function clean
algorithms which make errors in imaging extended structure when the
loop gain is much higher than 0.1, Multi-Scale CLEAN is able to
produce good images with a loop gain of 0.5 or even higher.

The algorithm can be made efficient by careful precomputation of all
terms that are needed for the update:

\begin{itemize}
\item $B * m(\alpha_q)$
\item $B * m(\alpha_q) * m(\alpha_r)$
\end{itemize}
By using these precomputed images, the operations needed for the CLEAN
reduce to scaling, shifting, and subtracting, just as for the \Hogbom\
algorithm. 

We now turn to the question of the shape of the components,
$m(x,y,\alpha)$. There are a number of considerations:
\begin{itemize}
\item Since the final image is composed of a summation of these components,
each individual component should be astrophysically plausible (for example,
without negative brightness.)
\item The component should be independent of pixel orientation, and
therefore should be a function of $\sqrt{x^2+y^2}$ only.
\item The component shape must allow a support constraint to be used. This
rules out Gaussians since the tails extend over all space. However,
too sharp a truncation in image space will cause difficulties as well.
\end{itemize}

For these reasons, we have chosen a tapered, truncated parabola as the
component shape:

\begin{displaymath}
m(r,\alpha) = \Psi(r){{\left(1 - {\left({r}\over{\alpha}\right)}^2\right)}}
\end{displaymath}
where $\Psi$ is a prolate spheroidal wave function (calculated using
an approximation provided by F. Schwab). The difference between this
function and a suitably scaled Gaussian is quite minor, though, and
use of a Gaussian will only be a problem at high dynamic range or when
using an image plane support constraint.

For zero $\alpha$, this corresponds to a Dirac delta function. We
always include such a scale size to ensure that the finest scales can
be fit.

The use of extended components as well as delta functions must be
accommodated in any support constraint that is used. If a window
function $w(x,y)$ is to be used then no emission in the model can leak
outside of this window. Hence each scale size $\alpha$ must be subject
to a trimmed window function $w(x,y,\alpha)$ such that a component
centered within the trimmed window function has no emission outside
the window function $w(x,y)$. The trimmed window functions are simple
to calculate. However, clearly if $\alpha$ is too large, a trimmed
window function may not exist. This simply means that a component of a
given size cannot be fitted into the specified window function
$w(x,y)$.

A convergence proof can be constructed following that given by
\citep{Schwarz1978}.  The most important limitation of Schwarz's
convergence proof is that the PSF should be positive-semidefinite. If
the PSFs are calculated accurately and the component shape is chosen
appropriately, then convergence is assured.

The algorithm is relatively low in complexity and can be implemented
straightforwardly by augmenting an existing CLEAN code. The only
mildly difficult part is using the precomputed cross terms correctly.

We have implemented this algorithm in the AIPS++ (now CASA)
Package. It is available in the {\tt deconvolver} and {\tt imager}
tools. AIPS++ also supports the following algorithms: \Hogbom\ and
Clark Clean, the Maximum Entropy Method (via the Cornwell-Evans
algorithm), and Multi-Resolution Clean (via a trivial script.) In
addition, AIPS++ supports an unpublished algorithm called Maximum
Emptiness developed by the author about twenty years ago.  This
algorithm is identical to MEM except that the entropy objective
function is replaced by a sech function. The motivation for this is to
approximate the minimization of the L1 norm. We include this algorithm
for comparison purposes.

Our Multi-Scale CLEAN algorithm is similar to the Multi-Resolution
Clean \citep{WakkerSchwarz1988}, the principal difference being that
our algorithm works simultaneously on all scales being considered
instead of sequentially. In other words, our algorithm is greedy only
with respect to component position and strength whereas
Multi-Resolution Clean is greedy with respect to component scale,
position, and strength. The main virtue of simultaneous searching is
that errors on selecting a scale size can be corrected immediately
rather than being frozen in. 

The representation used in Multi-Scale CLEAN is similar to that in the
Pixon method but the algorithms are quite different. The greedy
Multi-Scale CLEAN approach cannot claim to find a globally optimal
solution (as the Pixon method does). A global solution would be
superior (less bias and more compact) if it could be calculated for
Fourier synthesis problems.  Future work would do well to investigate
the possibility of adapting Compressive Sampling algorithms to this
particular domain and representation.

The choice of rotationally symmetric components is required to cut
down the dimension of the implicit search space - it is possible to
use representations such \cite{StarckEtAl2003} as curvelets to better
represent sharp edges but the overhead in searching would be very
substantial. This points to a clear shortcoming in our strategy of
explicit and exhaustive search in the component parameters.

In Multi-Scale CLEAN, the selection of scale sizes to evaluate over is
admittedly ill-defined but usually not too crucial. Too fine a range
such as an arithmetic progression wastes compute time, and too coarse
can lead to poor convergence. We have generally chosen a geometric
progress such as 0, 2, 4, 8, 16, 32 or 0, 3, 10, 30 pixels,
terminating at or below the largest scale expected. This is an area
that could be improved in future work.

Multi-Scale CLEAN is stable in the presence of a spatially varying
background provided sufficiently large scales are included in the
search. The large scale structure is then removed first, leaving the
fine scale structure to be estimated on an largely empty background.

\section{Demonstrations}

\subsection{Performance on real data}

To illustrate one of the key motivations for the Multi-Scale CLEAN
algorithm, we show an application to a dirty image of one spectral
channel of a galaxy (NGC1058) observed in HI emission with the Very
Large Array in D configuration. Since in this example, the
signal-to-noise ratio is not high and the source has very extended
structure, the fine-scale sidelobes are not as troublesome as the
broader sidelobes. In the dirty image (see figure 1), a broad negative
bowl is seen surrounding the emission. Although this may appear to be
a largely cosmetic defect, it does prevent accurate estimate of
integrated emission in cases such as this.

\begin{figure}
\plotone{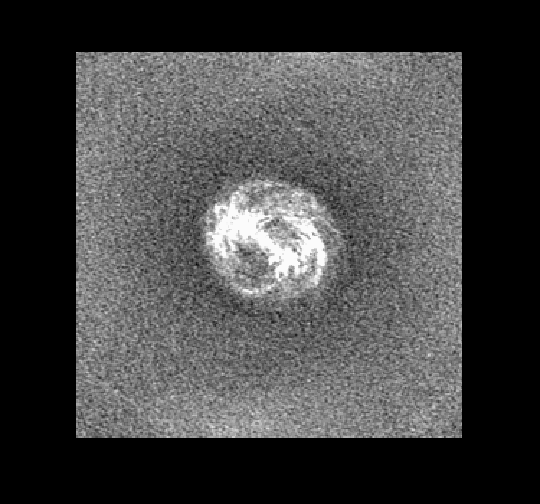}
\caption{Imaging of one channel 
of an HI synthesis of NGC1058, observed with the VLA. This shows the
``dirty image'' formed from inverse Fourier transform of the observed
data.  The transfer function is truncated to show the range -3mJy/beam
to +3mJy/beam. The peak in the image is about 12mJy/beam.}
\end{figure}

\begin{figure}
{
\plottwo{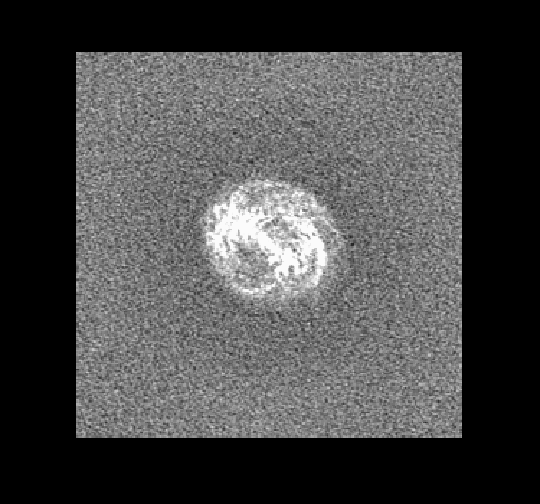}{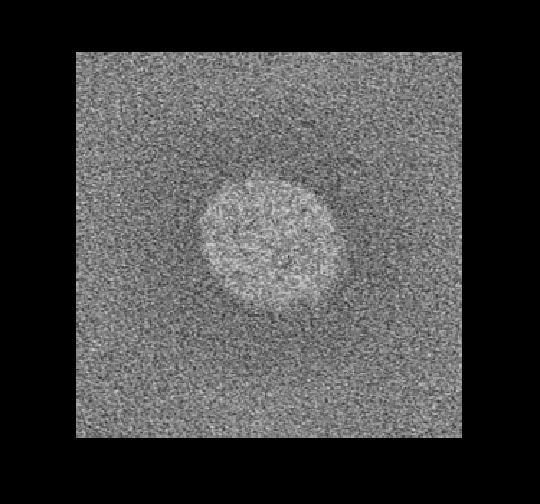}
}
{
\plottwo{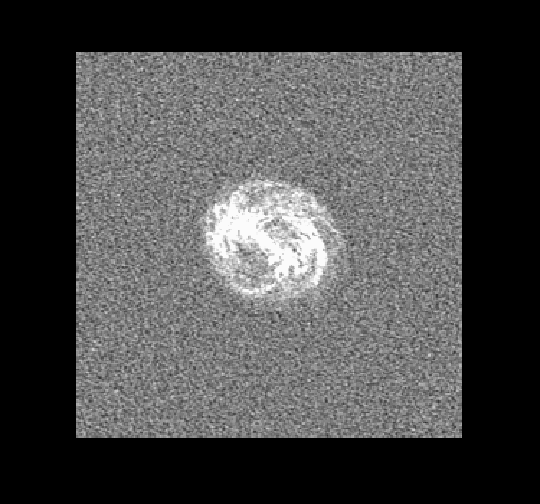}{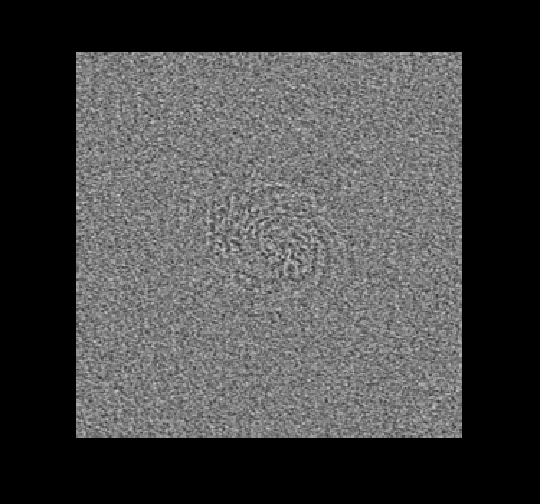}
}
\caption{Deconvolution at low signal-to-noise: imaging of one channel 
of an HI synthesis of NGC1058, observed with the VLA. 
(a) \Hogbom\ CLEAN restored image, 
(b) \Hogbom\ CLEAN residual image,
(c) Multi-Scale CLEAN restored image, 
(d) Multi-Scale CLEAN residual image. All images are displayed with the
same transfer function -3mJy/beam to +3mJy/beam. The peak in the image
is about 12mJy/beam.}

\end{figure}

Classic \Hogbom\ CLEAN (figure 2) is poor at correcting the negative
bowl because many point sources must be subtracted to represent the
broad emission. We applied Multi-Scale CLEAN to this image, using 1000
iterations at loop gain 1.0, with scale sizes set to 0, 1.5, 3, 6, 12,
24 pixels. The stopping criterion is that the peak residual flux reach
$4 \sigma$. At convergence, the integrated flux in the Multi-Scale
CLEAN image is 3.08Jy, and that in the \Hogbom\ CLEAN image is
0.32Jy. For the Multi-Scale image, this value is robust to changes in
the integration region to beyond the nominal extent of the source
whereas for \Hogbom\ CLEAN, the integrated flux is highly dependent on
the integration region (because of the substantial negative bowl in
emission left behind after deconvolution).

There are a number of features worth emphasizing:
\begin{itemize}
\item The Multi-Scale CLEAN algorithm finds and detects emission on the 
largest scales first, moving to finer and finer detail as iteration
proceeds. This is the opposite behavior to \Hogbom\ CLEAN where the fine
detail is removed first.
\item At large numbers of iterations for Multi-Scale CLEAN, the 
peak residuals for all scales become comparable.
\item The Multi-Scale CLEAN algorithm converges to relatively stable values of
the flux on different scales. In comparison, the \Hogbom\ CLEAN
continues to find more and more flux as iteration progresses.
\item To extract more flux using the \Hogbom\ CLEAN requires CLEANing
deeper than $4 \sigma$. This is possible, but is slow and distorts the
appearance of the noise.
\end{itemize}

\subsection{Performance on simulated data}

Now that we have seen how Multi-Scale CLEAN works on real data, we turn
to show behavior on simulated observations in which the ground truth is
known. 

First we show simulations of VLA observations on a well-known test
image - the ``M31'' image used in many radio simulation studies. This
is of a moderately complex source.  In figure 3, we show the model at
full resolution and also smoothed with a typical CLEAN beam (as would
occur in the usual restoration process).

Using the \AIPSpp\ {\tt simulator} tool, we simulated VLA observations
of this object in C configuration. The source was taken to be at
declination 45 degrees, and observations of 60 seconds were made every
5 minutes for hour angles from -4 to +4 hours. This is therefore a
well-sampled observation with sufficient short spacings that the
deconvolution should perform quite well.

\begin{figure}
\plotthree{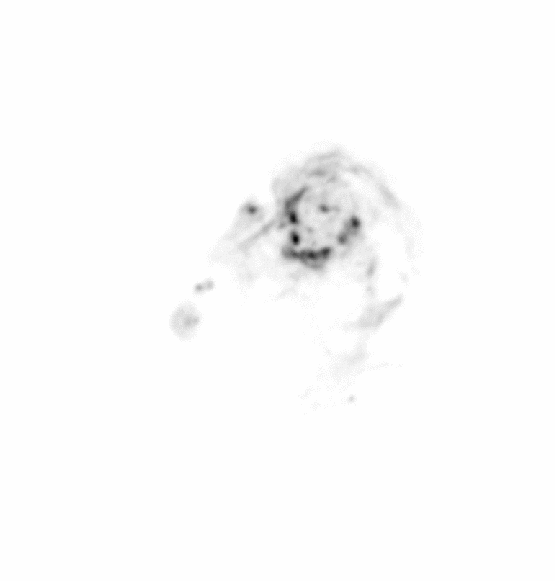}{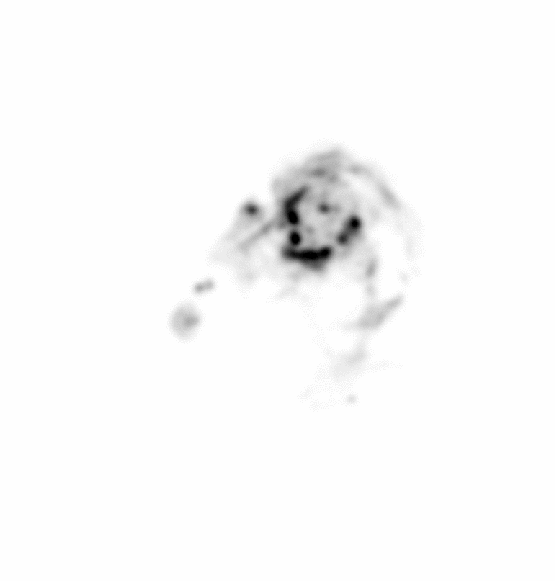}{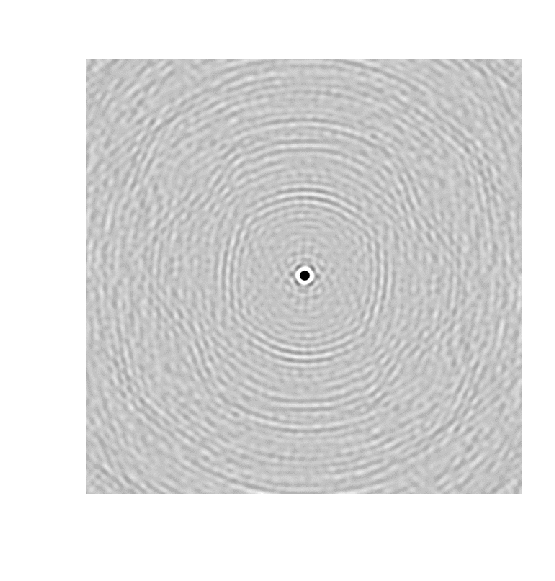}
\caption{The image and PSF used in simulations.
(a) Full resolution model image
(b) Smoothed model image.
(c) Point spread function.}
\end{figure}

We performed deconvolutions with the \Hogbom\ and Clark CLEAN,
and Maximum Entropy method, in addition to Multi-Scale CLEAN. The
Multi-Scale CLEAN used scales 0, 1.5, 3, 6, 12, and 24 pixels. In
figure 4, we show the various restored images, along with the error
pattern. The latter is calculated by subtracting the model image as
smoothed with the appropriate CLEAN beam.

\begin{figure}
\plotthree{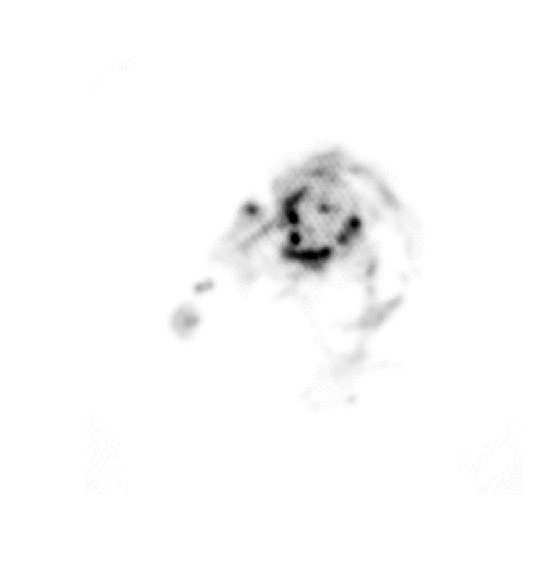}{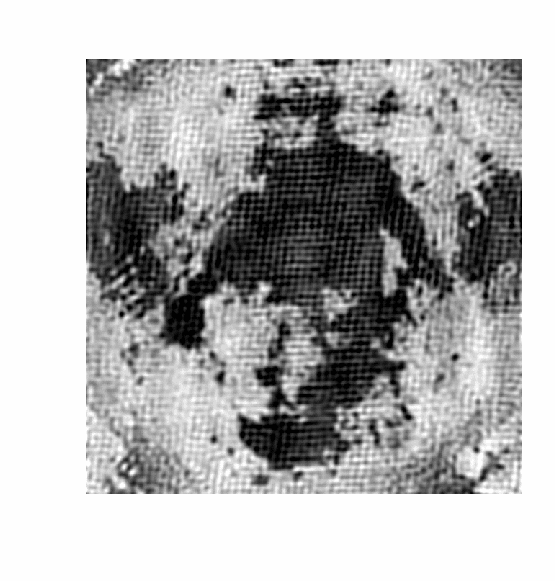}{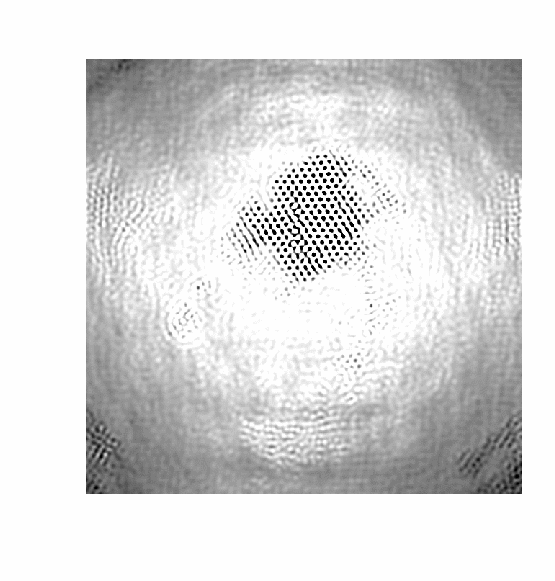}
\plotthree{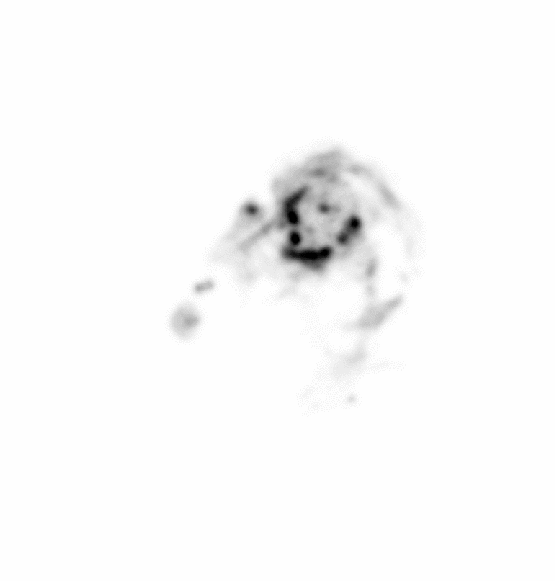}{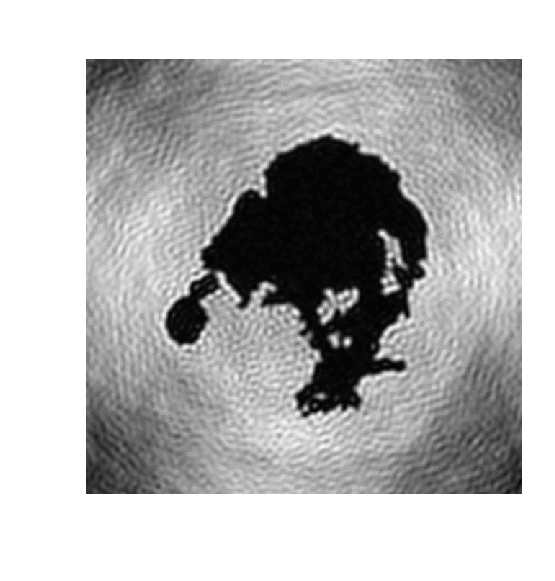}{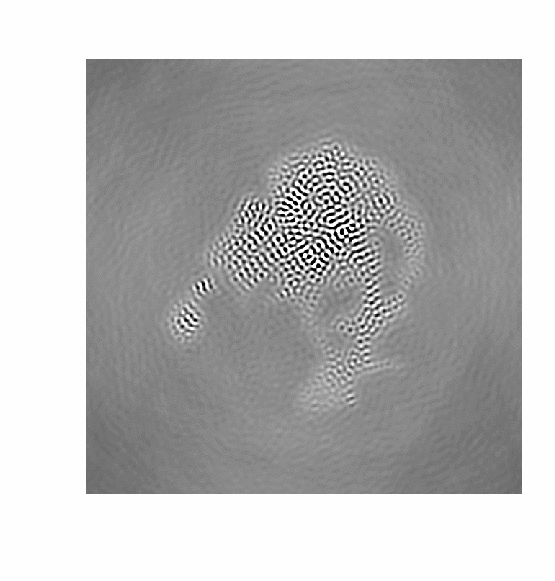}
\plotthree{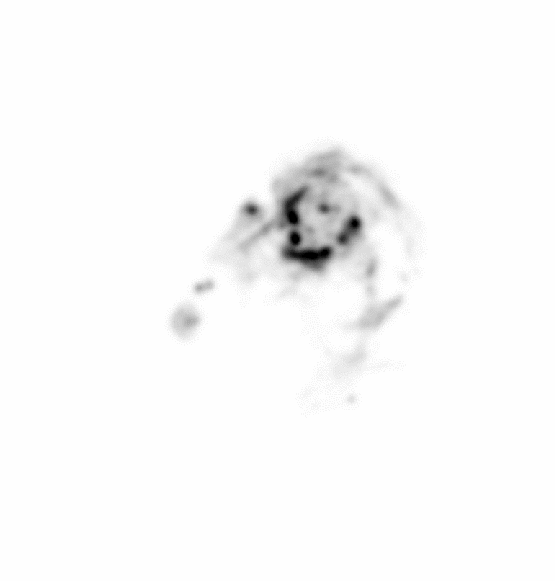}{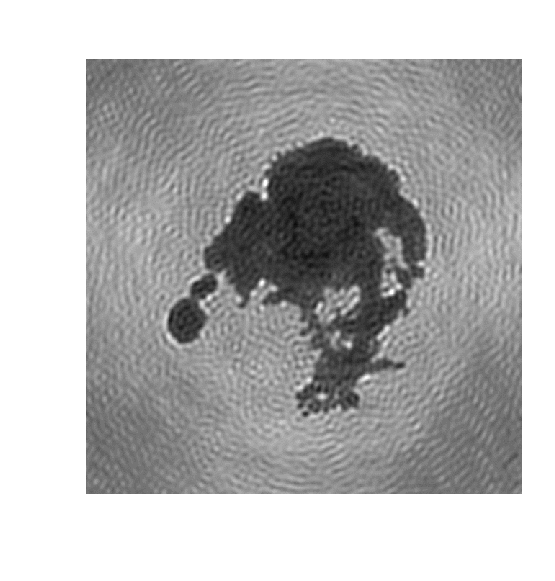}{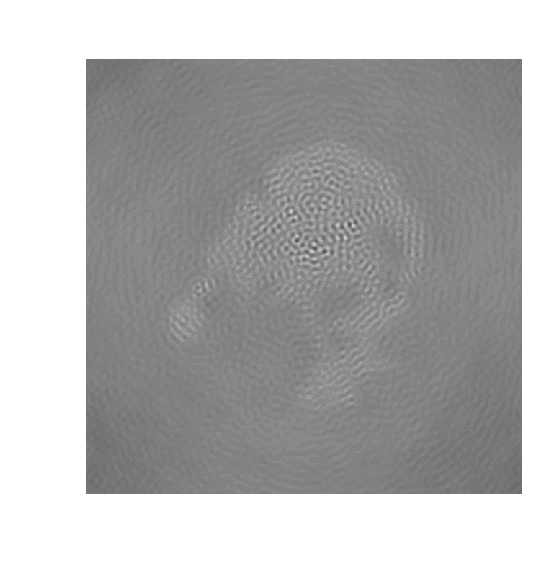}
\plotthree{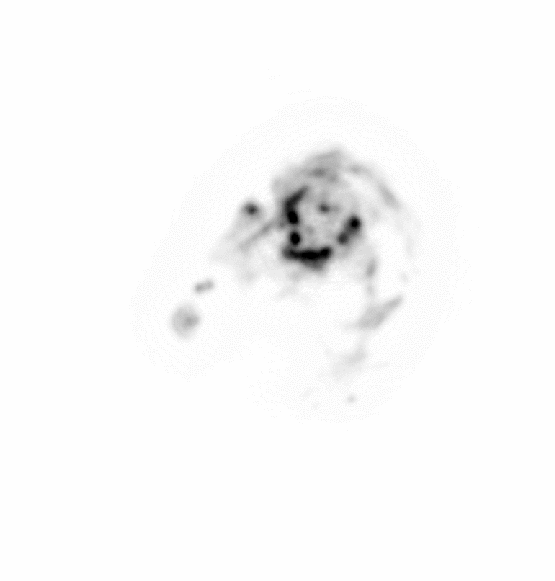}{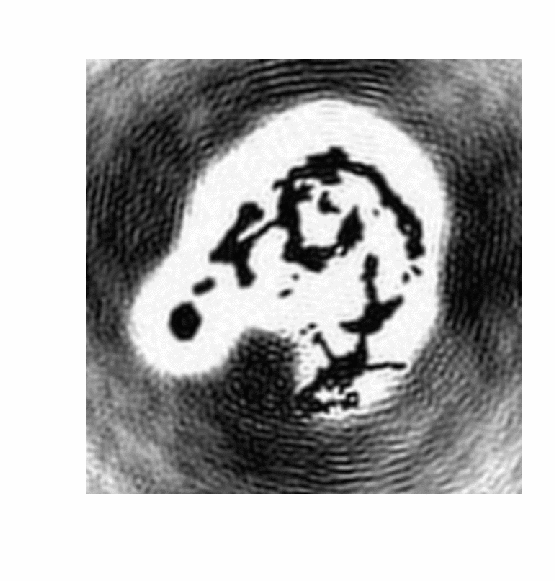}{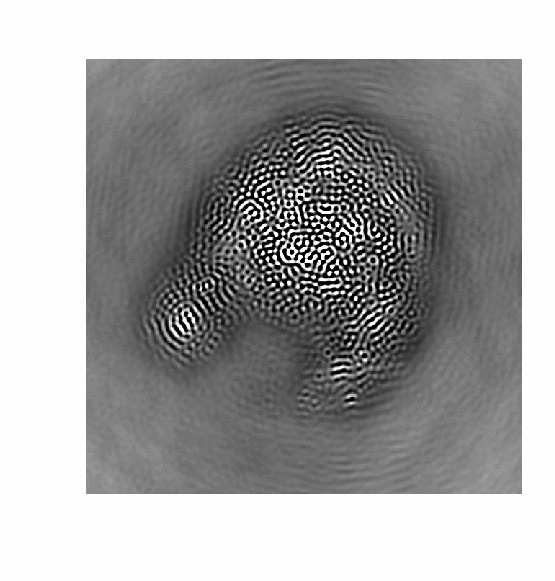}
\plotthree{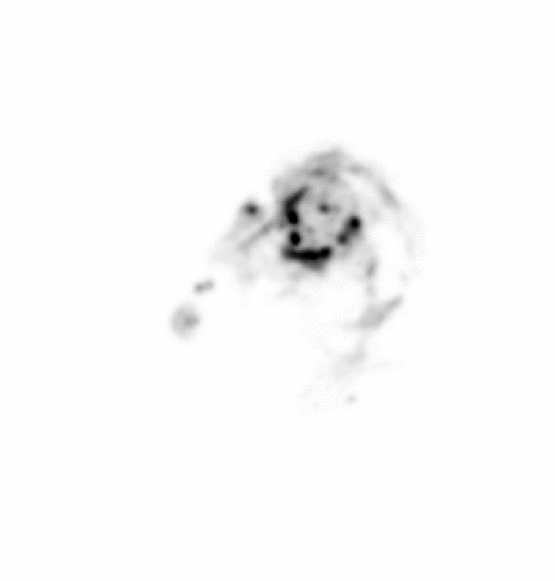}{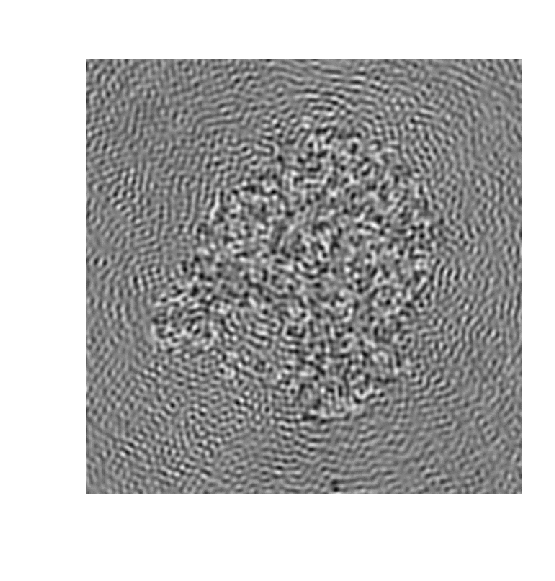}{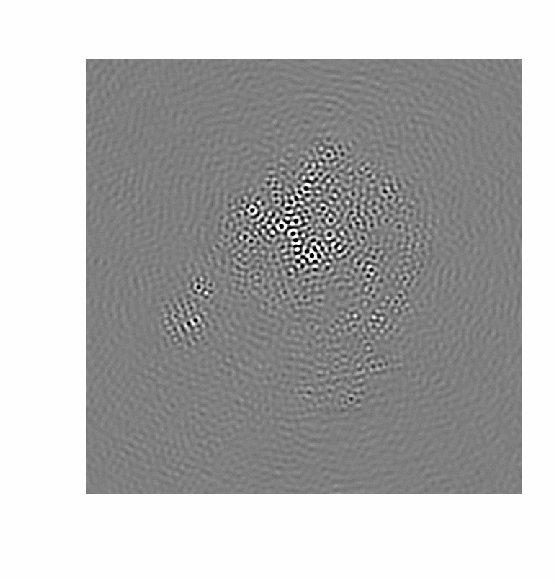}

\caption{Deconvolution of simulated VLA observations of the ``M31'' image.
(a) Clark CLEAN restored image, 
         (b) Clark CLEAN residual image, 
         (c) Clark CLEAN error image, 
         (d) \Hogbom\ CLEAN restored image, 
         (e) \Hogbom\ CLEAN residual image, 
         (f) \Hogbom\ CLEAN error image, 
         (g) Entropy restored image, 
         (h) Entropy residual image, 
         (i) Entropy error image, 
         (j) Multi-Resolution CLEAN restored image, 
         (k) Multi-Resolution CLEAN residual image, 
         (l) Multi-Resolution CLEAN error image,
         (m) Multi-Scale CLEAN restored image, 
         (n) Multi-Scale CLEAN residual image, 
         (o) Multi-Scale CLEAN error image. 
The restored images
are displayed with a transfer function running from -0.050 to 10Jy/beam,
the residual images from -1mJy/beam to +1mJy/beam, and the
the error images from -50mJy/beam to +50mJy/beam}
\end{figure}

The times taken for each deconvolution (on a 3.06GHz Xeon) are shown
in Table 2.  

\begin{table}
\caption{Performance of deconvolution algorithms for ``M31'' simulation.}
\begin{center}
\begin{tabular}{lcccc}
\hline
{Algorithm}&{Time (s)}&{Total flux (Jy)}&{RMS error (mJy/beam)}\\ \hline
\hline
\Hogbom\&204&1474&8.8\\
Clark&50&1326&25.6\\
Multi-Resolution&147&1546&14.7\\
Multi-Scale&614&1495&4.9\\
Entropy&13&1486&2.5\\
\hline
\end{tabular}
\end{center}
\end{table}

Some comments on the results:

\begin{itemize}
\item This simulation illustrated a difficult obstacle to the use of
  Clark CLEAN on deep images. Schwarz's \citep{Schwarz1978} analysis
  of the CLEAN algorithm shows that the PSF must be positive
  semi-definite (all eigenvalues non-negative). There are two reasons
  why the actual PSF may violate this requirement - first, the use of
  a gridded transform in place of a Fourier sum and second, for the
  Clark algorithm, the truncation of the PSF. Thus, at high iteration
  numbers, the CLEAN algorithm in either form may diverge. In this
  case, we found that the Clark CLEAN diverged above ~ 300,000 CLEAN
  components at which point only 1391Jy of the full 1495Jy were
  recovered. While some tuning of the details of the major/minor
  cycles might have helped, the point remains that Clark CLEAN is
  marginal at high iteration numbers. The \Hogbom\ CLEAN does converge
  well.  By comparison, the Multi-Scale CLEAN converges quite well at
  only 5000 iterations.
\item The structure in the residuals is quite obvious except for the
  case of Multi-Scale CLEAN, for which the residuals show little
  correlation with the source structure.
\item Entropy, \Hogbom\ CLEAN, and Multi-Scale CLEAN perform
  exceptionally well in recovering the full flux of the object,
  whereas the Clark CLEAN is biased down by about 7\%. This is in
  agreement with the qualitative results from the error images where
  the Clark CLEAN image shows a substantial negative bowl - exactly
  the effect Multi-Scale CLEAN was designed to avoid.
\item Even when the CLEAN algorithms converge, the results are prone
  to show ``fringing'' on extended emission
  \citep{Cornwell1983,BriggsCornwell1994,Briggs1995}. This fringing is
  due to poor interpolation in holes and at the edge of the sampled
  Fourier plane. The fringing is apparent in the error images
  (difference from ground truth) but not in the residual images. These
  simulations show that Multi-Scale CLEAN is also liable to such
  effects but at a lower level.
\item Entropy produces systematically biased residuals but the error
  is much lower than for the various CLEAN algorithms. In this case,
  the bias term causes an extra 2Jy of flux in the overall 1495Jy
  flux.  If thermal noise were added to the simulations, MEM's
  positivity bias would increase.
\end{itemize}

To test on a structurally different object, we have simulated VLA
observations of Hydra A. Our model is an actual image with noise
removed. The observing scheme was as in the M31 simulation. The
results are shown in figure 5.

\begin{figure}
\plotthree{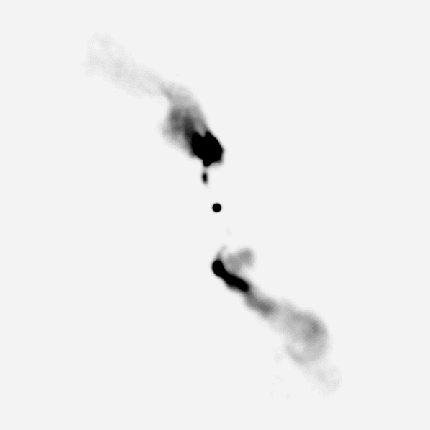}{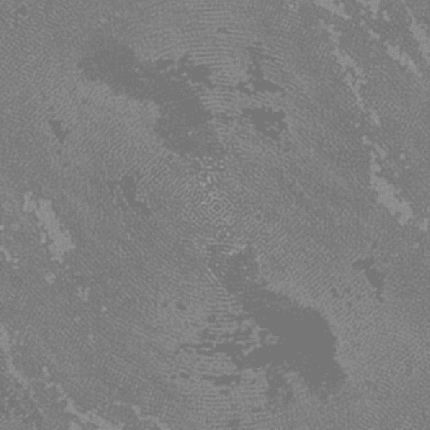}{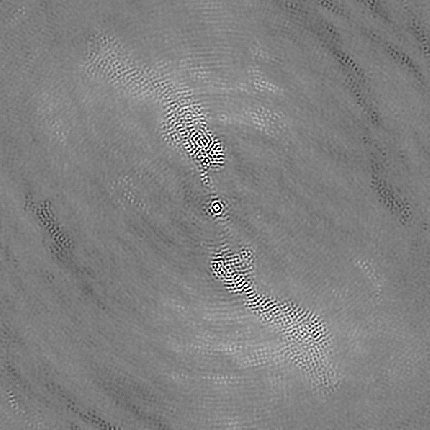}
\plotthree{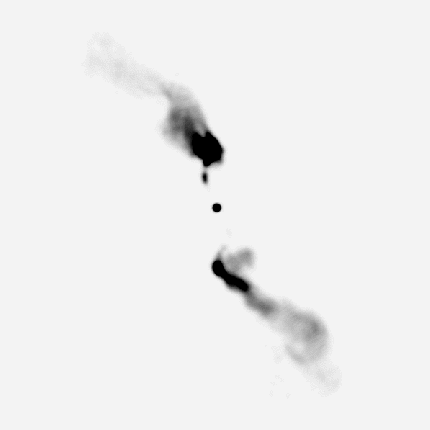}{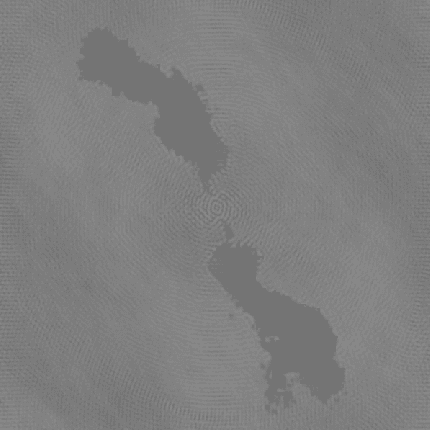}{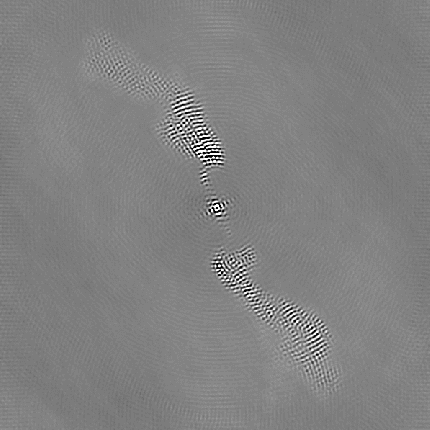}
\plotthree{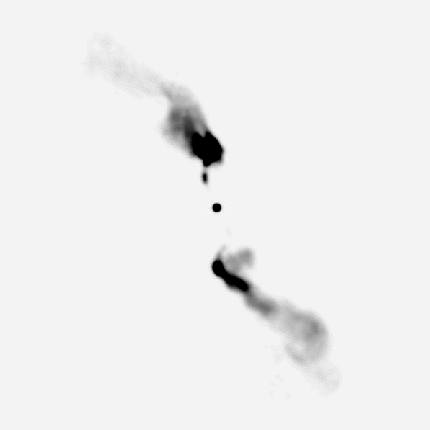}{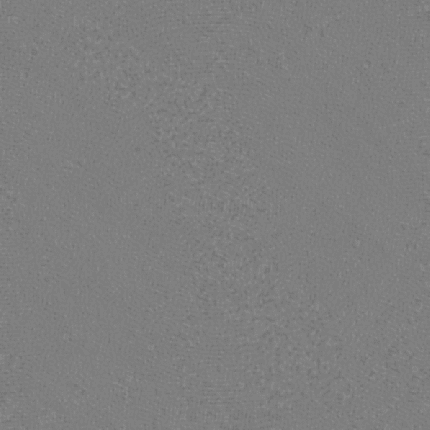}{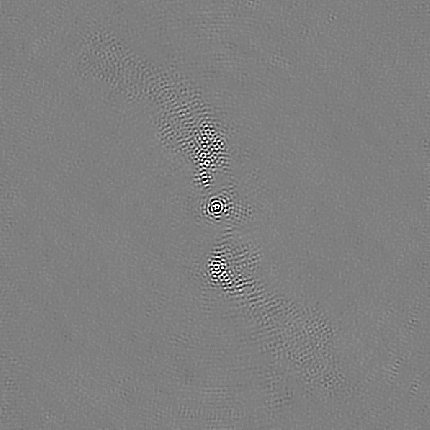}

\caption{Deconvolution of simulated VLA observations of the ``Hydra'' image.
(a) Clark CLEAN restored image, 
         (b) Clark CLEAN residual image, 
         (c) Clark CLEAN error image, 
         (d) \Hogbom\ CLEAN restored image, 
         (e) \Hogbom\ CLEAN residual image, 
         (f) \Hogbom\ CLEAN error image, 
         (g) Multi-Scale CLEAN restored image, 
         (h) Multi-Scale CLEAN residual image, 
         (i) Multi-Scale CLEAN error image. The restored images
are displayed with a transfer function running from -0.050 to 10Jy/beam,
the residual images from -1mJy/beam to +1mJy/beam, and the
the error images from -50mJy/beam to +50mJy/beam}
\end{figure}

\subsection{Performance for varying source size and noise level}

To gain a better understanding of the strengths and weaknesses of
Multi-Scale CLEAN algorithm and compared to other available
deconvolution algorithms, we have investigated the imaging numerically
through simulations made with systematically varying source size and
thermal noise.  These simulations were made with the \AIPSpp\ {\tt
  simulator} tool.  We put our model sources at declination 45~degrees
and observed them with the VLA C~array at X~band with 60~s
integrations between hour angles of -1 and +1.  For models, we used
both the ``M31'' brightness distribution and a model derived from an
optical image of the spiral galaxy M51.  Both images have been
modified so that the brightness distributions have well-defined edges
and off-source pixels are identically zero.  From each model, we have
produced a series of models with the same pixel size, but with the
linear size of the source distribution decreased in a geometric series
of $\sqrt{2}$.  To assess the image quality, we use two very simple
measures - recovered flux and dynamic range.  The integrated flux
itself can be an important quantity in many observations. In the
context of these simulations, deviations from the true flux indicate
systematic imaging errors which will in some way limit the usefulness
of the reconstructed image in quantitative endeavors.

\begin{figure}
\plotone{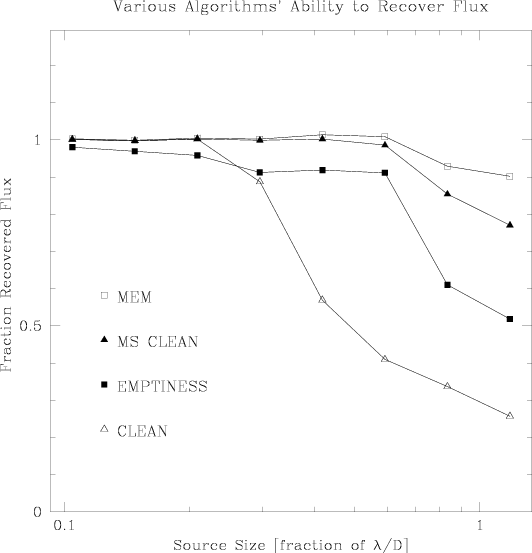}
\plotone{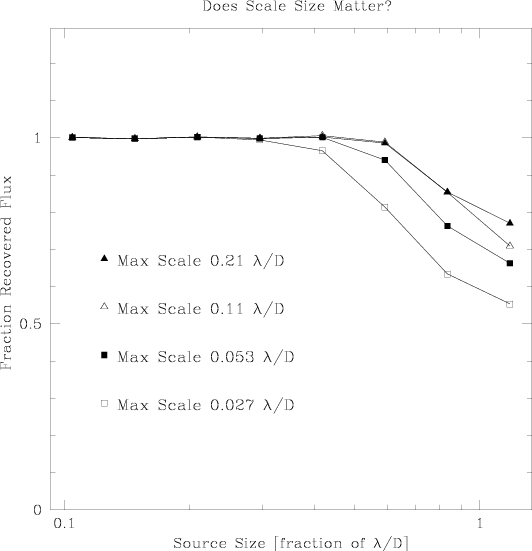}
\caption{The fraction of the recovered flux as a function of simulated
  source size (a) for various deconvolution algorithms, and (b) for
  Multi-Scale CLEAN with a variety of scale sizes.  In each case, four
  scale sizes were used, spanning from a point source to the maximum
  listed in the figure (which ranges from 8 pixels, or
  0.027~$\lambda/D$, to 64 pixels, of 0.21~$\lambda/D$). }
\end{figure}

Figure~6 indicates how the Multi-Scale CLEAN, Maximum Entropy Method
(MEM), Maximum Emptiness, and Clark CLEAN algorithms fail to recover
all the flux as the source structure becomes larger in the case of the
M51 series of model images.  It is well known that MEM is superior to
CLEAN at imaging extended structure. These simulations show this, but
also that Multi-Scale CLEAN is nearly as good as MEM in imaging very
extended structure.  Figure~6 also shows how the recovered flux depends
upon exactly which scale sizes are used in the Multi-Scale CLEAN
algorithm.  For smallish sources, the amount of flux recovered does
not depend upon the details of the scale sizes we used.  However, for
very large sources (the sources at about 0.8~$\lambda/D$ and
1~$\lambda/D$ are larger than we should be able to accurately image in
C~array), using large maximum scale sizes can appreciably increase the
amount of flux recovered.

\begin{figure}
\plotone{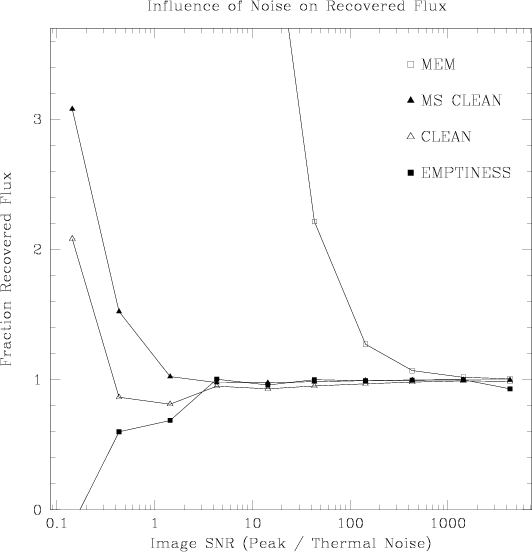}
\plotone{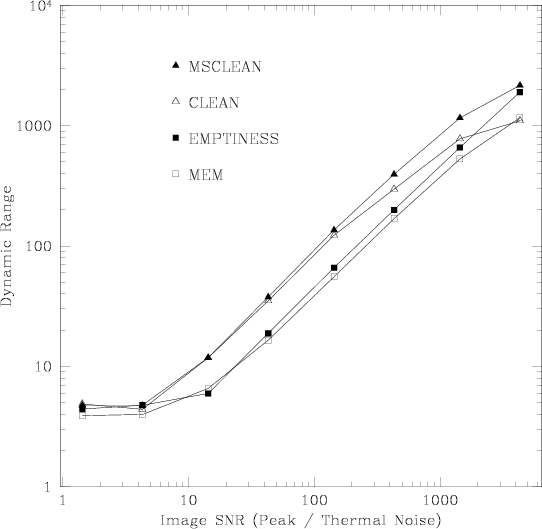}
\caption{The image quality as a function of image signal to noise ratio (a) fraction of recovered flux, and (b) dynamic range (defined as the reconstructed image
peak divided by the off-source RMS)}
\end{figure}

One of the major problems of the Maximum Entropy Method is positivity
bias.  Many extended sources which could benefit from MEM's superior
ability to image large-scale structure show weak diffuse structure at
moderate to low SNR.  Figure~7 shows the results of a set of
simulations done on the M51 model scaled to 0.30~$\lambda/D$ -- a size
which is small enough so that all algorithms do a fairly good job at
recovering all the flux.  Varying amounts of Gaussian noise are added
to the visibilities, and the theoretical image plane SNR, defined as
the peak of the model image convolved with the CLEAN beam divided by
the image plane noise, is calculated.  We gauge the deconvolution
algorithms' response to noise first by looking at the recovered flux
in Figure~7a.  Here, we see that MEM's positivity bias results in a
significant overestimate of the source flux starting with image plane
SNR as high as 100.  CLEAN, Maximum Emptiness, and Multi-Scale CLEAN,
do not share this shortcoming, and reproduce accurate estimates of the
source flux down to SNR of about 4.  Note that CLEAN actually
estimates the total flux reasonably well even when the image plane SNR
is below 1.  Of course, if we were to taper this image to lower
resolution, we would increase the amount of flux in the beam faster
than we increase the thermal noise.  In other words, \Hogbom\ CLEAN
and Multi-Scale CLEAN, are able to utilize the information in the
shorter baselines even when the source is totally undetected at full
resolution.  Figure~7b shows the dynamic range of the reconstructed
images as a function of the image plane SNR.  For SNR-limited imaging,
we would expect a straight line with slope 1.0.  At the high SNR end
of the plot, we are just beginning to see the curves flattening as the
images become limited by deconvolution errors.  The flattening at the
low SNR end of the plot at a dynamic range of about 4 is indicitave of
the peak over the rms of an image dominated by thermal noise.

\begin{figure}
\plotone{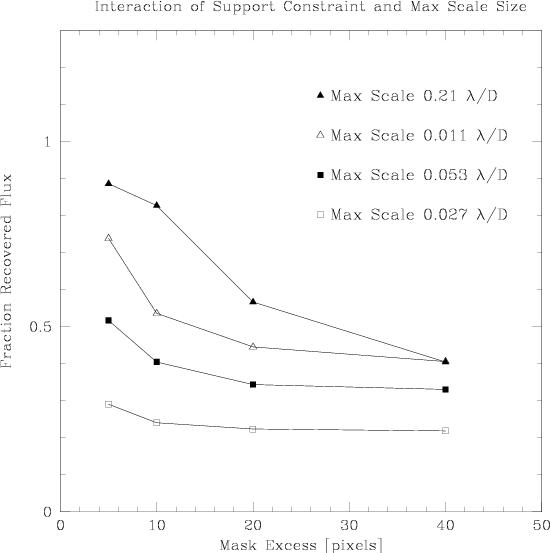}
\caption{Interaction between the maximum scale size
used in Multi-Scale CLEAN and the tightness of the mask used in the
deconvolution.  We used a model image
of the M31 HII region which has been scaled to a size of
0.83~$\lambda/D$.}
\end{figure}

Figure~8 points to the complex interactions between the maximum
scale size used in Multi-Scale CLEAN, the tightness of the imaging
mask, and the resulting image quality.  In this case, a large
model image (0.83~$\lambda/D$) was used, and Multi-Scale CLEAN was
just starting to fail.  As the plot indicates, significant improvement
in image quality can be obtained by using a fairly large maximum 
scale size (about one quarter the size of the object), and by using
a very tight mask image.
 
\section{Summary}

Multi-Scale CLEAN has been in use for 5-6 years.  It seems to work
well in practice. For examples of the application of the algorithm,
see \citep{Subrahmanyanetal2003, Momjianetal2003}.  Besides the obvious
efficacy of the algorithm, the major attraction is that it is simple
to understand and to implement. The most notable deficiencies are that
the scales must be chosen arbitrarily, and an arbitrary bias towards
small scales must be introduced. For extended emission, Multi-Scale CLEAN
is much faster than \Hogbom\ or Clark CLEAN but usually slower than MEM.
Multi-Scale CLEAN shows much less bias than MEM at low to moderate signal to
noise, and can be applied to images with negative flux.

\section* {Acknowledgement}

I thank Mark Holdaway for his contributions to this project.  I thank
my co-workers in the AIPS++ Project for providing the package within
which this work was carried out. I thank Sanjay Bhatnagar for many
stimulating conversations on multi-scale methods, and Rick Puetter and
Amos Yahil for discussions on pixon based deconvolution. In addition,
I thank Michael Rupen for providing the NGC1058 data.

I thank the referees of this paper for numerous helpful and illuminating 
comments and references.

\bibliographystyle{IEEETran}
\bibliography{msclean}

\end{document}